\documentclass{article}
\usepackage[utf8]{inputenc}
\usepackage{geometry}
\usepackage{multirow}
\usepackage{txfonts}
\usepackage{multicol}
\usepackage{booktabs}
\usepackage{authblk}
\usepackage{mathdots}
\usepackage{graphicx} 
\usepackage{fancyhdr}
\usepackage{hyperref}
\usepackage{float}
\usepackage{microtype}
\usepackage{biblatex}
\usepackage{caption}
\makeatletter
\def\@xfootnote[#1]{%
  \protected@xdef\@thefnmark{#1}%
  \@footnotemark\@footnotetext}
\makeatother

\usepackage{sectsty}
\subsectionfont{\normalfont\itshape}
\subsubsectionfont{\normalfont\itshape}
\title{\textbf{Detection and Classification of Brain tumors Using Deep Convolutional Neural Networks}}
\author[1]{Ranit Sen}
\affil[1]{Computer Science and Engineering}
\affil[1]{SRM Institute of Science and Technology, Chennai, India}
\affil[1]{rs6448@srmist.edu.in}
\author[2]{Gopinath Balaji}
\affil[2]{Computer Science and Engineering}
\affil[2]{SRM Institute of Science and Technology, Chennai, India}
\affil[2]{gb6580@srmist.edu.in}
\author[3]{Harsh Kirty}
\affil[3]{Computer Science and Engineering}
\affil[3]{SRM Institute of Science and Technology, Chennai, India}
\affil[3]{hk3829@srmist.edu.in}
\date{}
\begin{document}
\fancyhead{}
\fancyhead[C]{ }
\renewcommand{\headrulewidth}{0pt }
\fancyfoot{}
\fancyfoot[L]{\href{https://www.astesj.com}{www.astesj.com} \\ \href{https://dx.doi.org/10.25046/aj040501}{https://dx.doi.org/10.25046/aj040501}}
\fancyfoot[R]{\thepage}
\maketitle

\begin{multicols}{2}
\section{Abstract}
Abnormal development of tissues in the body as a result of swelling and morbid enlargement is known as a tumor. They are mainly classified as Benign and Malignant. Tumour in the brain is fatal as it may be cancerous, so it can feed on healthy cells nearby and keep increasing in size. This may affect the soft tissues, nerve cells, and small blood vessels in the brain. Hence there is a need to detect and classify them during the early stages with utmost precision. There are different sizes and locations of brain tumors which makes it difficult to understand their nature. The process of detection and classification of brain tumors can prove to be an onerous task even with advanced MRI (Magnetic Resonance Imaging) techniques due to the similarities between the healthy cells nearby and the tumor. In this paper, we have used Keras and Tensorflow to implement state-of-the-art Convolutional Neural Network (CNN) architectures, like EfficientNetB0, ResNet50, Xception, MobileNetV2, and VGG16, using Transfer Learning to detect and classify three types of brain tumors namely - Glioma, Meningioma, and Pituitary. The dataset we used consisted of 3264 2-D magnetic resonance images and 4 classes. Due to the small size of the dataset, various data augmentation techniques were used to increase the size of the dataset. Our proposed methodology not only consists of data augmentation, but also various image denoising techniques, skull stripping, cropping, and bias correction. In our proposed work EfficientNetB0 architecture performed the best giving an accuracy of 97.61\%. The aim of this paper is to differentiate between normal and abnormal pixels and also classify them with better accuracy.\\
\texttt{Keywords - Deep Learning, Convolutional Neural Network, Glioma, Meningioma, Pituitary, Transfer Learning}
\section{Introduction}
Cancer is an ailment caused by rapid and abnormal growth of cells in a specific region of the body which then spreads to other body parts. Cancer starts because genetic changes interfere with the orderly process of cell cycle where old cells die and new cells are formed. The abnormal cells together form a lump of tissue which is called tumor. A tumor finally gets classified as benign or cancerous. Tumor can be classified into two main types: benign and malignant. Benign tumors do not invade other sites of the body and remain in their primary location whereas malignant tumors invade other body parts and is often life threatening.\\
\indent The WHO (World Health Organization) states that cancer is the second leading cause of death worldwide. Moreover WHO also states that 9.6 million people have died worldwide out of which 30-50\% of the cases could have been prevented.  Even though the chances of developing Brain Tumor is approximately only 1\% according to Cancer Treatment Centers of America, in many situations it becomes quite challenging to diagnose the patient due to the similarity between, healthy and cancerous cells and also due to the similarity between different  types of brain tumor. Brain tumor is mainly categorized into three different types: Glioma, Meningioma and Pituitary tumors. The most common type of tumor found in the brain is Glioma. Meningioma tumors form in the meninges (a membrane that encompasses the brain). Pituitary tumors form in the pituitary gland which causes it to produce abnormal levels of hormones. A medical imaging technique called MRI (Magnetic Resonance Imaging) is used to detect tumors in the body.\\
\indent An MRI machine relies on radio waves and magnetic field to generate 2D cross-sectional images of a specific part of the body. The working of an MRI involves strong magnetic field that causes the protons in the body to align with the magnetic field. Then burst of radio waves are pulsed through the body causing the proton to move out of alignment. The protons move back to the correct position when the radio waves are turned off causing the release of energy which is detected by the sensors. Protons realign at different speeds because of different types of tissues in the body; this helps us differentiate between various tissues in the body. Even with such sophisticated techniques it is still quite difficult to manually distinguish between normal and abnormal cells. This is why Artificial Intelligence is often used to aid doctors.\\
\indent Due to the rapid developed in technologies like Deep Learning it has become much easier to segment, detect and classify brain tumors.  Convolutional neural network is one such area in deep learning that involves a lot of parameters, this make it very difficult to train a deep CNN model on an average laptop or desktop. Hence we have implemented famous CNN architectures using Transfer Learning, this enables us to use pre-trained weights and make minor changes to the model to suit our dataset.  We have also used various image denoising, data augmentation and segmentation techniques to improve the accuracy of the prediction, without any human intervention, for conventional Machine Learning classifiers as well as sophisticated Convolutional neural network model.
\section{Literature Survey}
Suhib Irsheidat, Rehab Duwairi \cite{irsheidat2020brain} uses simple ACNN architecture with 5 convolution layers for detection of Brain Tumors. Getting a validation accuracy of 88.25 \% which can be improved by performing bias correction and regularization along with Transfer Learning.\\
\indent S. Bauer, R. Wiest, L. P. Nolte and M. Reyes. \cite{bauer2013survey} with goal to provide a comprehensive overview of brain tumor and brain tumor imaging. \\
\indent Andac Hamamci et al. \cite{hamamci2011tumor} have presented techniques to segment tumors with that can assist researchers in planning for radiosurgery, and assessment for therapy effectiveness. They have implemented tumor-cut segmentation after the change in enhancing part.\\
\indent Jin Liu et al \cite{liu2014survey} studies of mind tumor division. He has talked about brain division techniques like limit-based division, fuzzy C Means, Atlas based division, Region division, Margo Random, etc.\\
\indent Sergio Pereira et al. \cite{pereira2016brain} introduced segmentation method on convolution neural network to identify tumor in brain. Automatic and reliable segmentation methods are introduced here. The main drawback of this paper is using automatic segmentation for classifying, the large spatial and structural variability among brain tumors may vary the results.\\
\indent Sobhaninia, Zahra et al \cite{sobhaninia2018brain} performed tumor segmentation using LinkNet. By using a single LinkNet network and using all seven training datasets, they successfully segmented the brain tumor image. The view angles of the images were not considered. They also introduced a method for CNN to automatically segment the most common types of a brain tumor which do not require preprocessing steps.\\
\indent Gopal, N. Nandha, and M. Karnan \cite{gopal2010diagnose} used Fuzzy C-Means clustering for brain tumor diagnosis and it was optimized with optimization algorithms. But it shows apriori specification of the number of clusters. The lower the value of beta we get the better result but a greater number of iterations are needed for the purpose.\\
\indent Mohsen, Heba, et al \cite{mohsen2018classification} uses 66 images and its goal is also to classify the brain tumors using DNN. The main drawback is that it used a very small set of data-set with no regularisation technique applied.\\
\indent Paper \cite{othman2011probabilistic} presents a Probabilistic Neural Network-based segmentation technique. Here they have Principal Component Analysis (PCA) for dimensionality reduction and feature extraction. Performance analysis is done with an accuracy of 73 to 100 percent with a standard deviation of +-13. So, the result can vary drastically.\\
\indent Devkota et al. \cite{devkota2018image} implemented methods to improve the computational time by Fuzzy C Means and Mathematical Morphological operation. Even though their proposed system detects cancer with 92\% accuracy and classifies with 86.6\% accuracy, they have not tested this up to the outcome and evaluation stage.\\
\indent \cite{abiwinanda2019brain} Implemented a simple architecture of CNN; with one layer each of convolution, flattening of layers, followed by a full connection from one hidden layer. Could use a better model by implementing Transfer Learning. Also, they haven’t compared their model results with any other model.
Gaurav Gupta et al \cite{gupta2017brain} proposed a new hybrid technique that is based traditional machine learning classifiers, like: Fuzzy C-Means clustering and Support Vector Machine, for brain tumor classification.\\
\indent In \cite{george2015brain} They used two classifiers depends on supervised techniques; the first classifier was the decision tree algorithm and the second classifier Multi-Layer Perceptron algorithm. Could have used some famous CNN architecture to compare results.\\
\indent Abhishek Anil \cite{anil2019brain} used Transfer Learning to classify Brain with tumor and Brain without tumor. Various well knows CNN architectures are used like VGG16, AlexNet, etc.\\
\indent Seetha, J and Raja, S Selvakumar in \cite{seetha2018brain} have performed brain tumor classification using CNN Classifier and got an accuracy of 97.5\%. Though they got better accuracy the model lacks preprocessing techniques like bias correction, skull stripping, denoising. So, when image with such characteristics are given the result will be affected.\\
\indent Ali Isin et al in \cite{icsin2016review} has focused on the recent trend of deep learning methods in this field and assessed the presented and future developments.\\
\indent Mazhar Shaikh in \cite{shaikh2019recurrent} has used a Recurrent attention mechanism-based network which aids in reducing computational overhead while performing Convolutional operations on highresolution images.\\
\indent Nikalal Kaldera et al. \cite{kaldera2019brain} have proposed a system that uses two different models; they have used CNN to classify the images and they have also used Faster Region based Convolutional Network to segment the images. This reduces the number of computations and achieves a high accuracy level.\\
\indent In \cite{deepak2019brain} their classification system used GoogLeNet to classify MRI images of brain tumor. GoogLeNet was implemented using Transfer Learning. But they have used only one model. Since they are implementing transfer learning they could have used multiple models and compared their results.\\
\indent Muhammad Sajjad et al. \cite{sajjad2019multi} have used InputCascadeCNN which is an automated deep learning segmentation technique. The architecture they have used is VGG-19 and to increase the size of the dataset they have also used data augmentation. They have used automated segmentation which can affect the result because of brain tumors spatial and structural variability.
\section{Proposed Methodology}
The proposed system involves MR Bias Correction and skull stripping to pre-process the images. Better segmentation by maximize the MRI image quality with minimized noise as brain MRI images very sensitive to noise. Gaussian blur filter, BM3D denoised, Total Variation smoothing was used in our work for noise reduction existing in Brain MRI which prevailed the performance of the segmentation. Along with these data augmentation is also used to increase the performance of the model and avoid overfitting. Using the concept of Transfer Learning to train the Convolutional Neural Network rather than using traditional approach to the problem. While in transfer learning, learning of new tasks relies on the previously learned task which has been trained over millions of images which not only makes the learning process faster but also makes it more accurate while predicting. Callbacks can aid in the faster resolution of errors and the development of better models. They may assist you in seeing how your model’s training is progressing, as well as preventing overfitting by introducing early stopping or changing the learning rate for each iteration. Here we will use TensorBoard, ModelCheckpoint, and ReduceLROnPlateau callback functions. For pooling methods we will be using GlobalAveragePooling2D instead of MaxPooling for creating second pooling layer while using transfer learning. Working architecture of the model is shown below in figure \ref{fig:MViTvHQi}.
\end{multicols}
\begin{figure}[H]
	\centering
	\includegraphics[width=12cm]{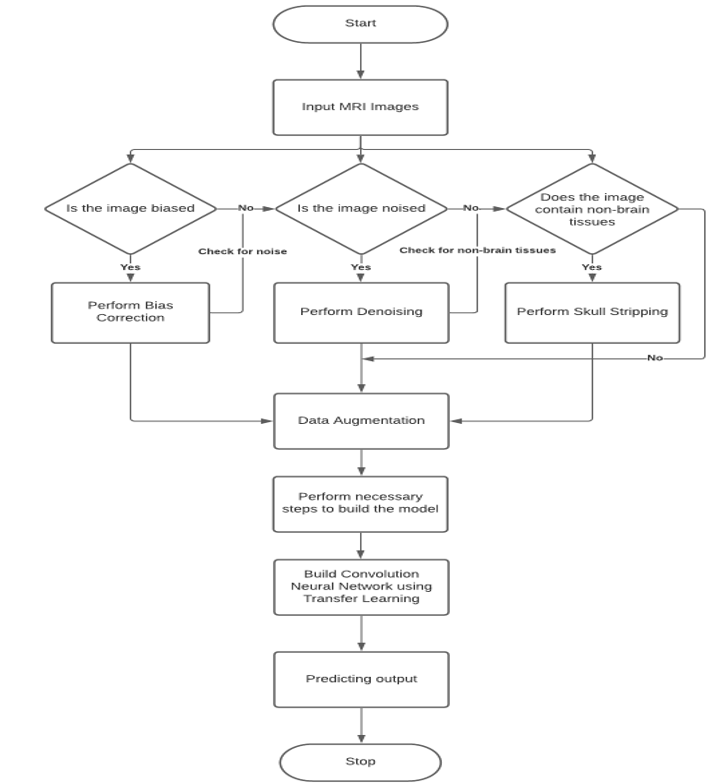}
	\caption{Working Architecture} 
	\label{fig:MViTvHQi}
\end{figure}
\begin{multicols}{2}
\subsection{Pre-processing of the image} 
\begin{figure}[H]
	\centering
	\includegraphics[width=9cm]{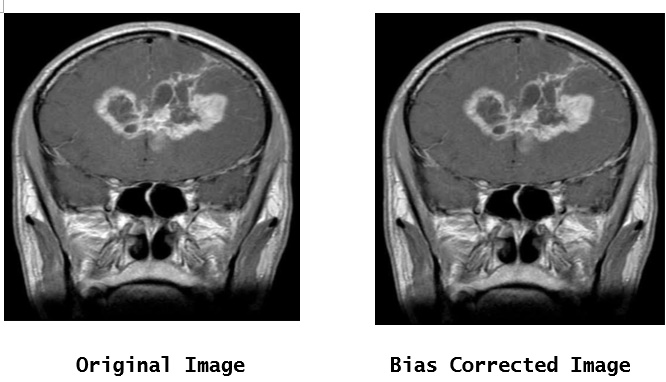}
	\caption{Bias Correction} 
	\label{fig:GBiTvHQi}
\end{figure}

\begin{figure}[H]
	\centering
	\includegraphics[width=9cm]{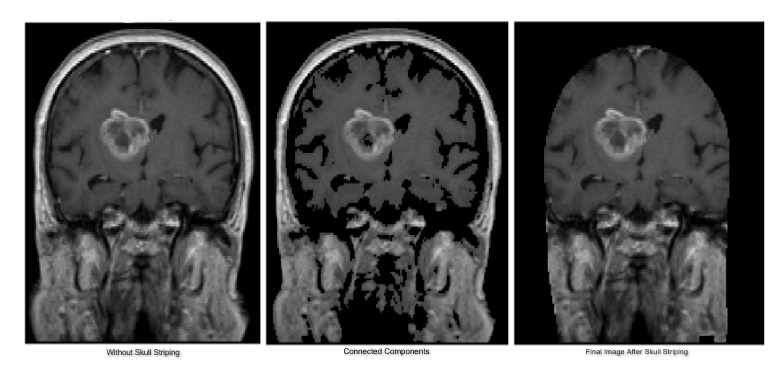}
	\caption{Skull Striping} 
	\label{fig:SBhSHiA}
\end{figure}

– 
1. Cropping and resizing the image - Before we applied any data pre-processing techniques, we first cropped and resized the images. Cropping the image allows us to remove the unnecessary black background that surrounds the MRI scans. Since it does not play any role in helping our model learning about the brain, it can be removed safely. Cropping is performed by finding the largest contour and then finding the extreme points of the largest contour. This following figure \ref{fig:DEBiTvHQi} shows the original image and the Figure Insert Number shows the image after cropping and resizing the image to the dimensions (150,150).
\begin{figure}[H]
	\centering
	\includegraphics[width=9cm]{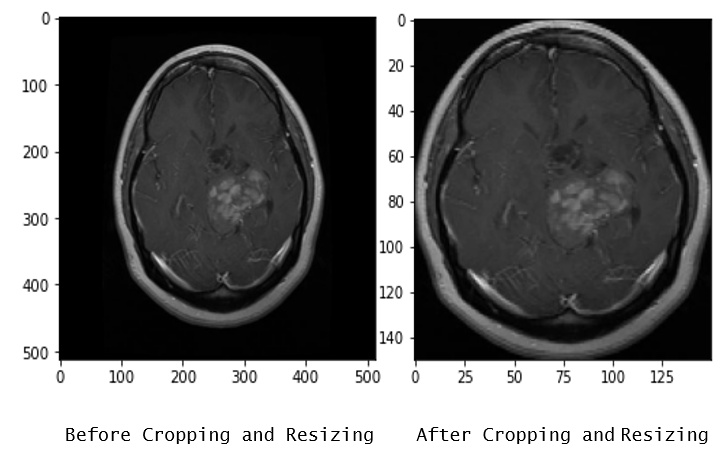}
	\caption{} 
	\label{fig:DEBiTvHQi}
\end{figure}
2.	Bias Correction - The bias field is a low-frequency spatially variable MRI artifact caused by magnetic field spatial non - uniformity, receiver coil sensitivity changes, and magnetic field interaction with the human body. For correcting this intensity non-uniformity of the MRI images we have used N4 Bias Field Correction. After performing the bias form the images are removed as shown in the figure \ref{fig:GBiTvHQi}.
3.	Noise Reduction – For denoising the image we have performed different noise reduction techniques mainly –\\
\indent a.	Gaussian Filter – Gaussian filter is a linear filtering process that is used to decrease the contrast and blur an image. By convolving the measured surface with a Gaussian weighting function, Gaussian filters may be applied to the input surface. A bell-shaped curve represents the Gaussian weighting function.
\begin{figure}[H]
	\centering
	\includegraphics[width=9cm]{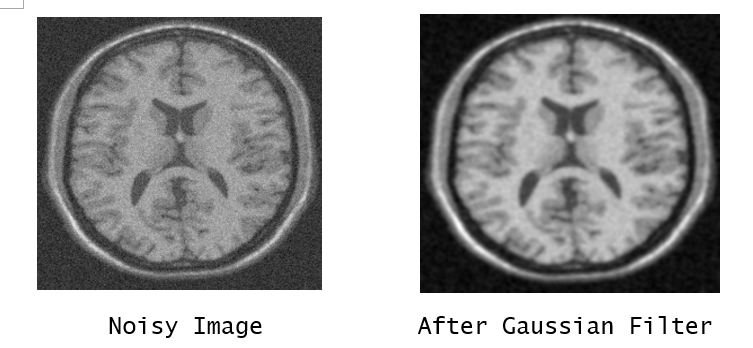}
	\caption{} 
	\label{fig:GcZFiTC}
\end{figure}
\indent b.	BM3D denoised – Block Matching and 3D filtering algorithm is a powerful tool for denoising the image with better results as compared to the other techniques. The suggested adaptive filtering technique's self-adaptation and stability have allowed it to achieve excellent noise reduction performance while preserving high spatial frequency information. We have used all the stages as most of the stage arguments have hard thresholding.
\begin{figure}[H]
	\centering
	\includegraphics[width=9cm]{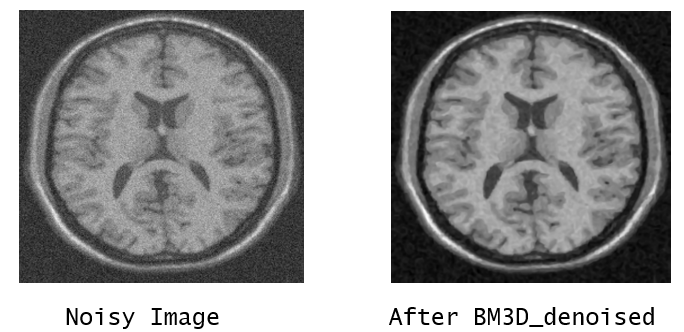}
	\caption{} 
	\label{fig:BcZFiTC}
\end{figure}
\indent c.	Total Variation - Total variation denoising (TVD) is a noise reduction technique designed to keep the underlying signal's sharp edges. A functional consisting of the sum of fidelity and regularisation components is minimised via TV regularisation.
\begin{figure}[H]
	\centering
	\includegraphics[width=9cm]{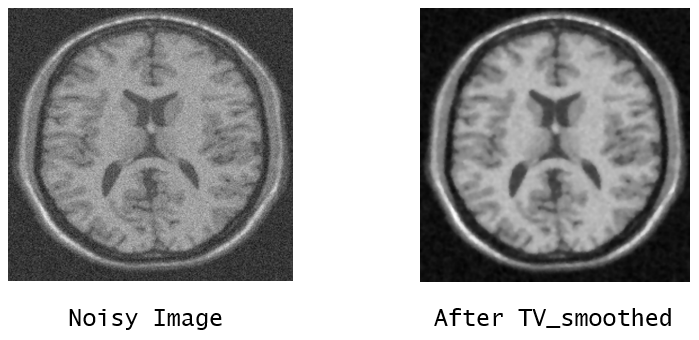}
	\caption{Kernel SVM} 
	\label{fig:BcZFiJC}
\end{figure}
\indent The best result which is obtained by BM3D denoising is used for further process.\\
4.	Skull Stripping – After removing the bias and the noise skull striping is performed. Skull stripping is an important step to pre-process the image. It is a process of removing the unwanted part in the MRI image like extra tissue, cortical surface, etc. To apply skull stripping we need to first check if the MRI scan is bimodal. This will help us verify if the image can be split into two different intensity classes. If the MRI scan is bimodal, we can proceed to apply Otsu’s method of image thresholding which will give us a binary mask of the brain. Using the threshold, the brain can be extracted by finding the largest connected component. But sometimes when the input image is noisy the extracted brain image might contain many holes in it. Which is why we also need to use a Closing Transformation to close the holes and extract the brain properly. Following \ref{fig:SBhSHiA}a shows the input image, \ref{fig:SBhSHiA}b shows an attempt to extract the brain without applying closing transformation and \ref{fig:SBhSHiA}c shows the final image after applying Closing Transformation. As visible from the final image, Skull Stripping has allowed us to reduce noise considerably.
5.	Data augmentation – This technique is often used to increase the performance of the any models by adding more data. Data augmentation may be utilised to satisfy both, the amount of data as well as variety of the training data. Data augmentation can also be used to rectify class imbalances when modelling for a classification task. Data augmentation techniques used are rotation, horizontal flipping, vertical flipping, width shift and height shift.
\subsection{Proposed Methodology for CNN}
\begin{figure}[H]
	\centering
	\includegraphics[width=9cm]{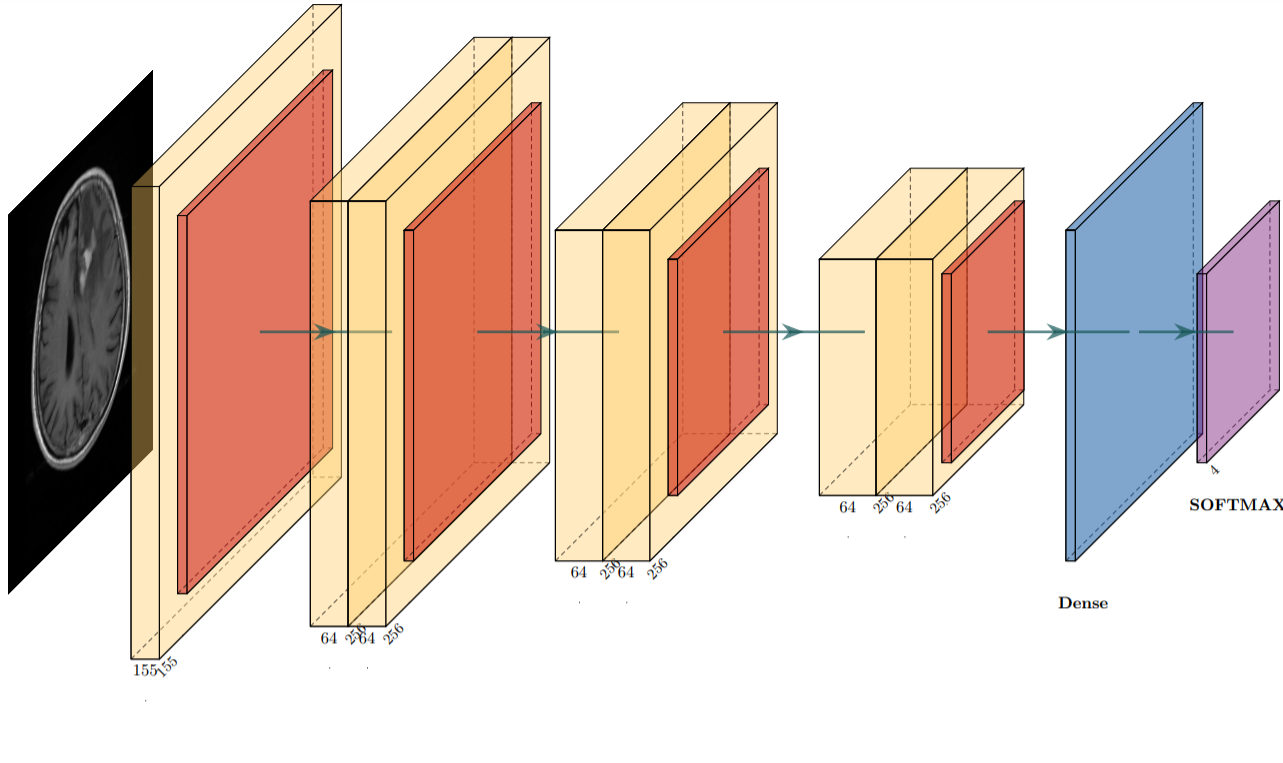}
	\caption{Convolution Neural Network} 
	\label{fig:SBhSHia}
\end{figure}

\begin{figure}[H]
	\centering
	\includegraphics[width=9cm]{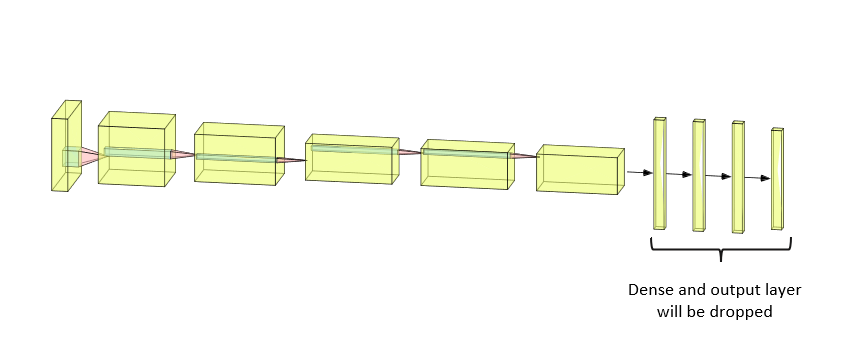}
	\caption{Transfer Learning with Convolution Neural Network} 
	\label{fig:SBMhSHia}
\end{figure}

Convolutional Neural Network (CNN) if figure \ref{fig:SBhSHia} is one of the most famous type of Artificial Neural Network that is mainly used for object detection, image classification and image segmentation. Due to the rapid developments in Artificial Intelligence, CNN proved to be an amazing tool for medical image-oriented tasks. We chose to use a CNN architecture over any other model because CNN employs several convolutional filters to scan the whole feature matrix and perform dimensionality reduction, hence making them well suited for image processing and classification. Famous CNN architectures like VGG16 \cite{simonyan2014very}, MobileNet \cite{howard2017mobilenets}, Xception \cite{chollet2017xception}, ResNet \cite{he2016deep} and EfficientNet-B0 \cite{tan2019efficientnet}. All of these architectures were implemented using Transfer Learning. Architecture of Transfer Learning is shown in figure \ref{fig:SBMhSHia}. Features like Dropout layer, Batch Normalization, Categorical Cross Entropy loss function, Adam optimizer, Global Average Pooling, and ReduceLROnPlateau are used to prevent overfitting and improve accuracy of the model.\\
\indent ReduceLROnPlateau is a very helpful function offered by Keras. It reduces the learning rate when a particular metric of choice has stopped improving. Keras claims that models are improved when the learning rate has been reduced by a factor of 2 to 10 once learning stagnates. If the model does not improve for a pre-set ‘patience’ number of iterations then the function will reduce the learning rate. For our model the patience value is set to 2.
\subsubsection{Classification Framework}
To correctly detect the edges of the MRI scans we employed customized and refined state of the art CNN models. The edges or features are effectively detected towards the top of our mode, i.e., close to the Fully Connected (FC) layers. A softmax function in the last classification layer will categorise data into 4 different tumor classes.\\
\indent The last layers of all the models were removed to adapt our model to our target domain. So, the FC layers were removed and new FC layers were added with output of 4 because we have to classify between 4 classes: glioma tumour, meningioma tumour, pituitary tumour and no tumour. The softmax layer was also replaced. Categorical cross entropy loss function was used because we are dealing with a multi-class classification problem.\\
\indent To replace the Dense layers, we added 1 Global Average Pooling layer followed by a Dropout layer. The Global Average Pooling is used instead of the traditional Flatten layer because the number of parameters is lesser, this will make the computation a little faster.\\
\indent The initial learning rate of the model is set to 0.0003. A very low value is selected compared to the default of 0.01 set by Keras because a high learning rate may overshoot and not converge. We used a Batch Size of 32, which is a generally accepted value in the Machine learning community, and then trained the model for 12 epochs. Too many epochs will cause the model to overfit, similarly the model will not be optimal if the number of epochs is too small.
\subsubsection{Model Settings}
\textit{Inductive Transfer Learning}:  Transfer Learning is a process by which a model trained on a specific task is reused and modified for a similar task. A formal definition given by Sebastian Ruder [30] is: If we have a source domain denoted by Ds, source task Ts, target domain Dt and target task Tt, our objective is to enable us to learn target conditional probability distribution $P(Yt|Xt)$ in Dt with the information gained from Ds and Ts where $Ds != Dt$ or $Ts != Tt$. Even under Transfer Learning there are multiple techniques which depend on the target and source domains. The method implemented in this research is called Inductive Transfer Learning. This method of Transfer Learning is used when the target and source labelled data is available in the target and source domains for a classification. The models, where transfer learning was implemented, are briefly explained in the following statements.\\
\indent VGG16 is a CNN model that was proposed in the year 2014 by Karen Simonyan and Andrew Zisserman of the Visual Geometry Group Lab of Oxford University \cite{simonyan2014very}. The ‘16’ in its name alludes to the fact that it contains 16 weighted layers. It was the first runner up of ILSVR (Imagenet large scale visual recognition challenge) in the year 2014, where the task was to classify 1.2 million natural images in Imagenet into 1000 defined classes. What makes VGG16 stand out is that the padding and maxpool layers in the model are always the same with 2x2 filter and stride of 2. The hyperparameters are also less because the filters used are of dimension 3x3 with a stride of 1. It uses a SoftMax function for the output, this comes after 2 fully connected layers. Due to the fact that VGG16 is a deep CNN architecture containing 13 convolutional layers and 5 max pooling layers, the number of parameters it contains is in the neighbourhood of 138 million.\\
\indent Xception is a CNN model that was proposed in the year 2016 by Francois Chollet \cite{chollet2017xception}. From the name of the architecture, it evident that is it an adaption of Inception model \cite{simonyan2014very}. It is termed as Xception because it implements the ideas of Inception architecture but to an ‘eXtreme’ level. Even though it has only 23 million parameters, which is almost equal to the number of parameters in Inception-v1, it outperforms most other architectures like Inception-v3 \cite{szegedy2016rethinking} when applied to the ImageNet dataset. It consists of 36 convolutional layers and 71 layers in total. What makes this model unique is that it has entirely used depthwise separable convolutions instead of Inception modules. This means 1x1 convolutions are used to capture the cross-channel correlations for every channel and 3x3 convolutions are used to capture the spatial correlations for each output. The original depthwise separable convolution layer consists of depthwise convolution which is then followed by a pointwise convolution, but the Xception architecture has modified this by keeping the pointwise convolution first which is followed by depthwise convolution.\\
\indent Our third model for brain tumour classification was ResNet, created by Kaiming He et al. \cite{he2016deep} in the year 2015. This is one of the most unique CNN architectures because of the introduction of the Residual Block. A Residual Block consists of some layers and a ‘skip connection’ with no parameters. The job of a skip connection is to just add the output from the previous layer to the layer ahead. This skip connection in ResNet has attenuated the problems of training a very deep network which are vanishing and exploding gradient. To avoid the issue of different dimensions in the skip connection, a convolution layer can be used to control the dimensions of the output volume.  The implementation of the Residual Block allows ResNet to go much deeper than previously possible. This ultimately helped them win the ILSVRC classification competition in the year 2015.\\
\indent MobileNet is a CNN class that was open-sourced by Google \cite{howard2017mobilenets}, and it provides us with an ideal starting point for training our ultra-small and ultra-fast classifiers. It employs depthwise separable convolutions, which means that instead of combining all three and flattening them, it executes a separate convolution on each colour channel. The channel-wise $D_{K} * D_{K}$ spatial convolution is called depthwise convolution. If we have five channels, we will have 5 $D_{K} * D_{K}$ spatial convolutions. The 1*1 convolution to modify the dimension is pointwise convolution. The key difference between MobileNet and typical CNN architecture is that instead of a single 3x3 convolution layer, the batch norm, and ReLU are used. The convolution was divided into a 3x3 depth-wise convolution and a 1x1 pointwise convolution using Mobile Nets.\\
\indent The last model we used was EfficientNet-B0 which was introduced by Mingxing Tan and Quoc V. Le \cite{tan2019efficientnet} in the year 2019. This model servers as a base for the family of EfficientNet models that range from EfficientNet-B0 to EfficientNet-B7. The EfficientNet family of models is probably the most advance CNN architecture because it achieves high accuracy while using a considerably smaller number of parameters. The EfficientNet-B0 is a mobile architecture that only has 11 million trainable parameters. This model uses 7 inverted residual blocks and Swish activation function (it is a multiplication of Linear and Sigmoid activation). The reason why these models perform so well is because it does not solely focus on scaling for depth like the ResNet models. Here width, resolution and depth are all scaled in a compound manner leading to better results.\\
\subsubsection{Dropout}
In CNN models, Dropout \cite{srivastava2014dropout} is method to prevent overfitting. Overfitting is a situation when the model fits the training data too well due to which the generalization of the model gets affected. So, the model may not perform well on data that isn’t a part of the training set.\\
\indent Dropout is a mechanism that randomly sets the input units to 0 at each iteration of the training set. Hence this has the effect of nullifying the contributing of some neurons by temporarily removing or dropping it from the network. To balance the sum of all the inputs, those that are not set to 0 are scaled up by 1/(1-rate).\\ \indent Dropout takes a value between 0 and 1, this signifies the fraction of the inputs units to be dropped. The dropout value for our models is 0.2. This means that there is a 20\% chance that the output of a certain neuron will be driven to 0.  
\subsubsection{Global Average Pooling}
Pooling, in CNN models, is a layer added after the convolution layer that down samples the spatial size of the image. This is down to ‘summarize’ the image and reduce the number of parameters and hence also reducing the computation cost. Pooling layer operates by sliding a filter over the image. At each window either the average or maximum value is taken to pass on to the next layer.\\
\indent Global Average Pooling \cite{lin2013network} performs a similar operation to average pooling, that is it takes the average value of each window of the filter. The only difference is that Global Average Pooling computes a single average for each channel.\\
\indent The advantage of Global Average Pooling is that the number of trainable parameters reduces significantly. This in turn will also reduce the chances of overfitting. The authors of this paper, Min Lin et al., also state that removing a fully connected layer for a Global Average Pooling forces the feature maps to be more closely related to the classification categories.
\subsubsection{Batch Normalization}
Batch Normalization \cite{ioffe2015batch} is a regularization method that prevent the model from overfitting. Normalization is a data standardisation method used in pre-processing. This enables us to use different sources of data inside the same range. The data is normalized to have a mean of 0 and standard deviation of 1.\\
\begin{equation}
\begin{array}{l}
x_{norm} = \frac{X-m}{s}
\end{array}
\end{equation}
Where the $ x_{norm} $ is the normalized data point, X is the data point to normalize, m is the mean and s is the standard deviation.\\
Batch Normalization is not applied on the data directly, instead it is a normalizing technique that is applied between layers of the network. It is defined as:\\
\begin{equation}
\begin{array}{l}
z^N = \frac{z-m_{z}}{s_{z}}
\end{array}
\end{equation}
Where $ m_{z} $ is the mean of the neurons, $ s_{z}$ is the standard deviation of the neurons’ output.
\subsubsection{Categorical Cross Entropy Loss Function}
A loss function is used to calculate and minimize the error of a model during the optimization process. It is used when the model weights need to be adjusted during training. The amount of loss is computed depending on how far it deviates from the actual value. Since the aim is to minimize the loss, smaller loss usually means better model. There are various options to choose from when selecting loss function. The loss function selected for this model was Categorical Cross Entropy which is also called logarithmic loss.\\
\indent Categorical Cross Entropy is used for multi-class classification model. The penalty for this loss is logarithmic in nature, producing a high score for big deviations near to 1 and a low score for tiny discrepancies close to 0. So, a Cross Entropy of 0 means the model in perfect. The loss is defined by:\\
\begin{equation}
\begin{array}{l}
Loss = -\sum_{i=1}^{output-size} yi\log \hat{y_{i}}
\end{array}
\end{equation}
\subsubsection{Adam Optimizer}
Adam Optimizer Optimization in Machine Learning is the process of iteratively training the model to provide a maximum or minimal function evaluation of the Optimization Algorithm. The Optimization Algorithm used for this research is Adam \cite{kingma2014adam}.\\
\indent Adam derives its name from adaptive moment estimation and combines the finest aspects of the AdaGrad and RMSProp optimizers. Adam can be used in place of Stochastic Gradient Descent (SGD) to update weights of the network. It is an adaptive moment estimation method because it adjusts the learning rate for each weight of the neural network by estimating the first and second moments of gradient. It is the most common optimizer because it is computationally efficient, easy to implement and works well for problems that involve large amounts of data or parameters.
\section{Dataset Description}
The Kaggle dataset is freely accessible and is frequently used for testing classification algorithms. The dataset consists of 3264 .jpeg images of MRI scans of 3 types of brain tumours: meningioma, glioma, and pituitary tumours. It also consists images of MRI scans of brain with no tumour. The MRI images in this dataset are from 3 different views: sagittal, coronal, and axial views. The dataset consists of 500 MRI scans of brain with no tumor, 901 MRI scans of pituitary tumor, 937 images of meningioma tumor and 926 MRI images of glioma tumor. All the images are resized to dimensions 150x150. Since the number of MRI scans with no tumor is much lower, it was separately augmented to increase the number of images. This will also help make number of images of all classes approximately the same and prevent the model from classifying an image to the class in which there are more training samples.\\
\section{System Limitations}
There are certain restrictions in this article since ML approaches are utilised for detection and classification of brain tumors. Any machine learning model is known to learn from previous data or data presented to ML models; nevertheless, the accuracy of forecasting accurate outcomes might vary noticeably each time they predict an output and may occasionally offer different results if provided a different type of image for which they are not trained.\\
	\indent And, because machine learning approaches rely heavily on data, if the dataset used to train the model is small, there may be a lot of volatility, which might impair the test set or development set performance. As a result, a model's performance cannot be inspected or guaranteed over time. This problem can be overcomed by continuously providing the image when we get them from any occasion.\\
	\indent Another drawback of ML is that it is stochastic rather than deterministic. It does not take into account the science behind MRI or X-Ray image or how it is measured. It simply learns from the information given to it. So, any uniformity in the system can affect the result. But we have handled it to some extent by performing bias correction, noise reduction and skull stripping.
\section{Environment}
For this project we have used Python and have implemented that in jupyter Notebook. Python is an object oriented, high-level, interpreted programming language. Jupyter Notebook (open-source code) is a development environment for creating and executing Python code that originated as the iPython Notebook project. The python libraries used mainly for this project are numpy, pandas, matplotlib, tensorflow, openCV, skimage, nipype, keras, seaborn, scipy and scikitlearn. The experiments were performed using Jupyter Notebooks on a device with 8GB RAM Intel i5 10th Generation processor coupled with Nvidia GTX 1650 Graphic Card.
\section{Experiment Result}
\subsection{Evaluation and Metrics}
\begin{figure}[H]
	\centering
	\includegraphics[width=9cm]{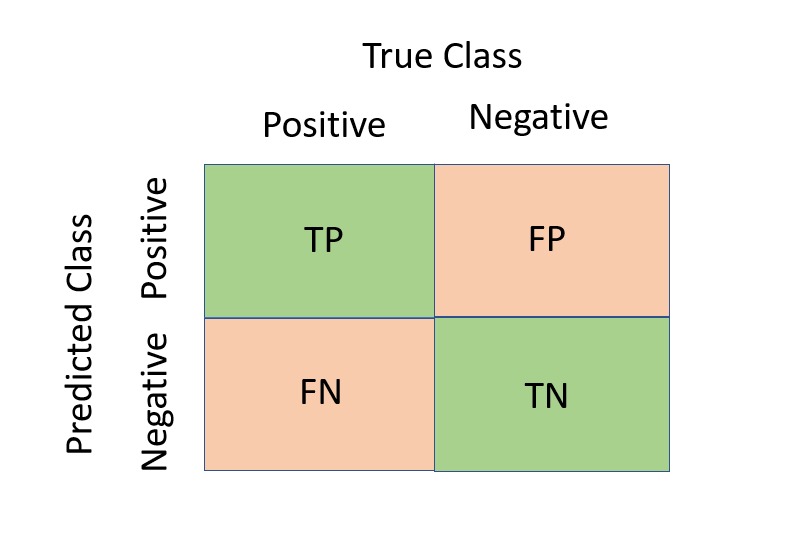}
	\caption{Confusion Matrix} 
	\label{fig:BNrPvQi}
\end{figure}

\begin{figure}[H]
	\centering
	\includegraphics[width=9cm]{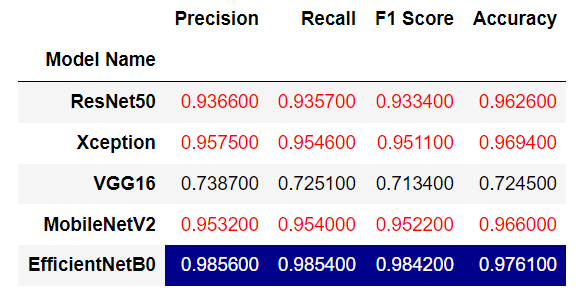}
	\caption{Comparsion Table} 
	\label{fig:BNrKvQi}
\end{figure}

\begin{figure}[H]
	\centering
	\includegraphics[width=9cm]{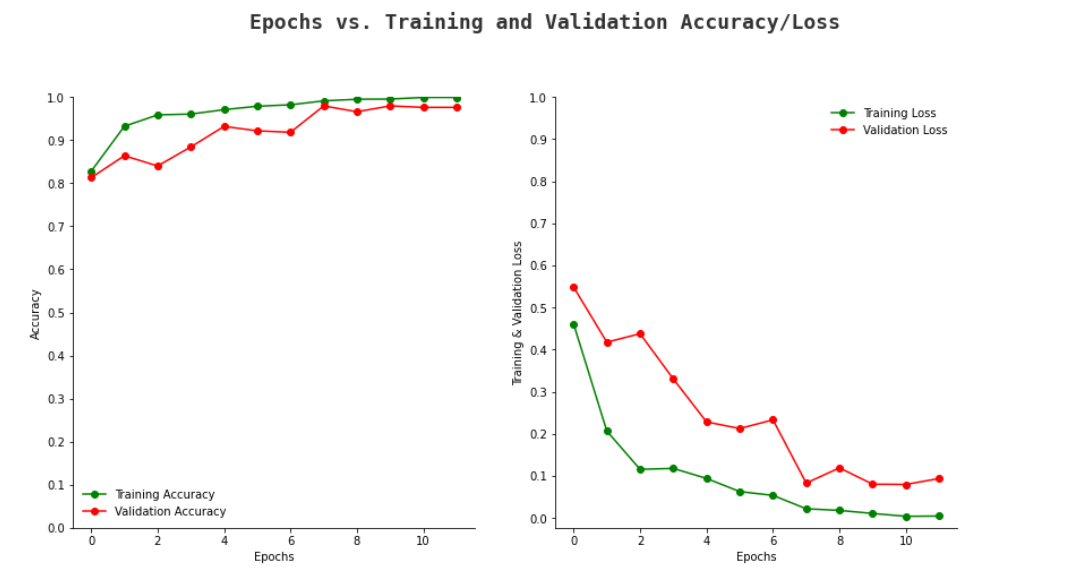}
	\caption{Accuracy and Loss Curve} 
	\label{fig:BNrMvQi}
\end{figure}

\begin{figure}[H]
	\centering
	\includegraphics[width=9cm]{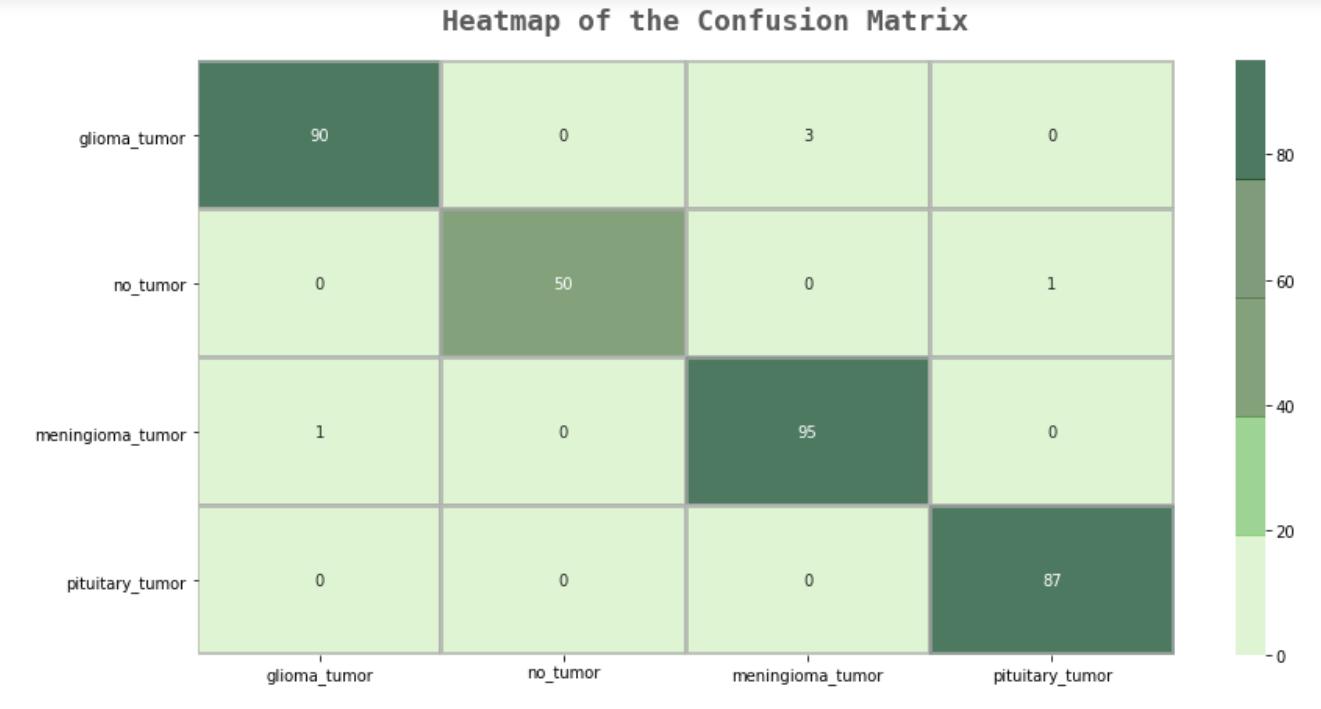}
	\caption{Confusion Matrix of the Model} 
	\label{fig:LNrMvJi}
\end{figure}

\begin{figure}[H]
	\centering
	\includegraphics[width=9cm]{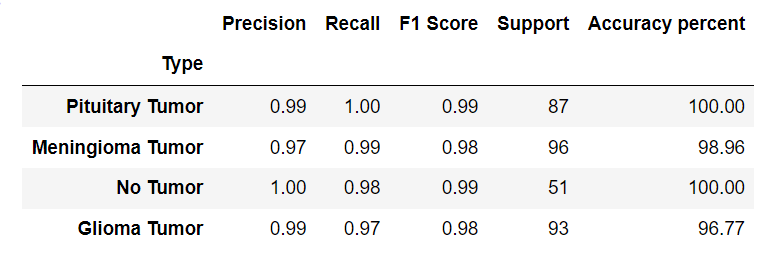}
	\caption{Results} 
	\label{fig:LNrMvQi}
\end{figure}

For the standard assessment of a classifier, several performance metrics are established. The metrics used to evaluate our models are Accuracy, Specificity, F-score, Precision and Recall.\\
\indent For starters, for the classification algorithms, we employed a confusion matrix and drew conclusions from it. A confusion matrix is a table that provides information about a model's performance or quality for two or more types of classes given a set of test data for which the real values are known. A two-dimensional confusion matrix is the most basic confusion matrix for classifier 2 class shown in firgure \ref{fig:BNrPvQi}.
•	True Positives (TP) - A true positive result is one in which the model accurately predicts the positive class.\\
•	True Negatives (TN) -Likewise, a true negative result is where the model accurately predicts the negative class.\\
•	False Positives (FP) - This is a result where the model incorrectly predicts the positive class.\\
•	False Negatives (FN) - Similarly, this is an outcome where the model incorrectly predicts the negative class.\\
From the confusion matrix we can predict accuracy, precision, recall and F1-score.\\
\indent •Accuracy - Accuracy is a statistic used to describe the performance of the model in general across all classes. It is determined by dividing the number of correct predictions by the total number of predictions.\\
\begin{equation}
\begin{array}{l}
Accuracy = \frac{TP+TN}{TP+FP+FN+TN}\\
\end{array}
\end{equation}
\indent •Specificity: This is the ratio of the actual negatives that were predicted as negative, i.e., true negatives.\\
\begin{equation}
\begin{array}{l}
Accuracy = \frac{TN}{TN+FP}\\
\end{array}
\end{equation}
\indent •Precision: It is defined as the ratio of, number of true positives by the total number of positive predictions.  It answers the question ‘what portion of positive predictions were actually correct’.\\
\begin{equation}
\begin{array}{l}
Precision = \frac{TP}{TP+FP}\\
\end{array}
\end{equation}
\indent •Recall: It is defined as the ratio of the number of true positives by the total number of true positives and false negatives. It answers the question ‘what portion of actual positives were identified correctly’.\\
\begin{equation}
\begin{array}{l}
Recall = \frac{TP}{TP+FN}\\
\end{array}
\end{equation}
\indent •F1 Score: It is defined as the weighted average or harmonic mean of Precision and Recall.
\begin{equation}
\begin{array}{l}
F1 = \frac{2 * (Precision * Recall)}{Precision + Recall}\\
\end{array}
\end{equation}
From \ref{fig:BNrKvQi}, we can see that out of the 5 architectures used, EfficientNetB0 gave the best result which is 97.61\% in terms of accuracy. It was expected to perform the best considering it is the most recent and one of the most advanced architectures. Apart from EfficientNetB0, three other architectures gave a great accuracy of around 96\%: ResNet50, Xception, and MobileNetV2. Despite having the least number of parameters, MobileNetV2 giving a high accuracy of 96.60\% was unexpected. Despite repeated hyperparameter tuning VGG16 was unable to detect brain tumours properly. This issue could be attributed to the dataset being small and containing MRI scan from 3 different views (sagittal, coronal, and axial), so the model has very few images of each view and class to learn form. Since VGG16 does not employ any skip connections there may be a lot of information loss as we go deeper into the network. Hence this may be a reason as to why it does not perform well on this dataset.\\
\indent By gradually adding fresh training examples, learning curves plot the training and validation loss of a sample of training examples. We can use learning curves to evaluate if adding more training instances might enhance the validation score (score on unseen data). If a model is overfit, adding more training examples may help the model perform better on unknown data. Similarly, adding training instances won't assist if a model is underfit. There are mainly two types of curves we will be using accuracy curve and loss curve. Figure \ref{fig:BNrMvQi} shows the accuracy and loss curve. Accuracy curve depicts the accuracy of the model per epochs. We have used 12 epochs for this experiment. Loss curve does the same job but it calculates the loss.
The confusion matrix is drawn for the given model shown in figure \ref{fig:LNrMvJi}.\\
\indent When we compare other metrics too, it is evident that from EfficientNetB0 we got the best results in terms of Precision, Recall and F1 Score. \ref{fig:LNrMvQi} shows the performance metrics for each class of tumour predicted using EfficientNetB0. The prediction accuracy for the No tumour and Pituitary tumour class is calculated to be 100\%. Meningioma and Glioma prediction accuracy is not far behind with 98.96\% and 96.77\% respectively.
\section{Conclusion and Future Scope}
This research presents various data pre-processing techniques like noise reduction using skull stripping and BM3D, bias correction and data augmentation that allow us classify brain tumours with high accuracy. We have also came to a conclusion after running 5 different transfer learning models hence it is also providing a comparative study about different techniques also. Our proposed system applies state of the art CNN architectures using the concept of Transfer Learning to extract features from the images. Since the performance of our classification models were evaluated by reliable metrics, it helps certify the soundness of our proposed system. We also found a way work to work with very little data as this is usually the case in most real-world scenarios.\\
\indent Even though, we achieved high accuracy there are various improvements possible. Firstly, we have applied skull stripping on the whole data in one go. This may be a fast process but it doesn’t necessarily generate the best images. Second, we have only worked with CNN models that classify images. An improvement would be to detect the actual location of the tumour with bounding boxes using segmentation and sophisticated detection algorithms like YOLO (You Only Look Once) and SSD (Single-Shot Detector). Another modification that is possible is to use a traditional classifier, like K-Means, instead of using Dense layers and softmax for classification. Future research in this domain can lead to better results, possibly using better pre-processing methods and further fine tuning the model hyperparameters.

\end{multicols}
\printbibliography
\end{document}